\documentclass{aipproc}
\layoutstyle{6x9}

\newcommand{\be}{\begin{eqnarray}}
\newcommand{\ee}{\end{eqnarray}}
\newcommand{\bc}{\begin{center}}
\newcommand{\ec}{\end{center}}
\newcommand{\fr}{\frac}

\begin{document}

\title {Heat transfer in  theory of relativity}
\author{\footnote{brotas@fisica.ist.utl.pt} A. Brotas}{
  address = {Departamento de F\'{\i}sica, Instituto Superior T\'ecnico, \\
             Av Rovisco Pais, 1096 Lisboa Codex, Portugal}}
\author{\footnote{joao.carlos@tagus.ist.utl.pt}  \underline{J.C. Fernandes}}{
  address = {Departamento de F\'{\i}sica, Instituto Superior T\'ecnico, \\
             Av Rovisco Pais, 1096 Lisboa Codex, Portugal}}


\begin{abstract} 
The traditional Fourier equation just allows us to study the evolution of temperature in an "undeformable" bar.
The search for its relativistic variant is a task which is expected to fail because in relativity there are no undeformable bars.
Rigid bodies, in the sense of "as rigid as possible", are deformables.
In this work we show how to write in relativity the system of equations necessary to study simultaneously 
deformation and temperature evolution along a rigid deformable bar. The solutions of the two simultaneous equations is discussed
assuming convenient constitutive relations for the material. An application is presented.

\end{abstract}

\maketitle

\section{Introduction}
Since the beginning of relativity, some physicists have been busy looking for an alternative form of the
Fourier equation that excludes the possibility of energy and signal transmission at a velocity
larger than $c$. We think that this will be a task with no success.

In classical physics the Fourier equation is the heat transmission equation in an undeformable bar.
However, in relativity rigid and undeformable are not synonymous. The undeformable body is a geometrical concept.
The rigid body, in the sense of "as rigid as" possible (the body in which shock waves propagate with
velocity $c$) is deformable. Heat flux carries momentum and energy. 
So, in relativity, the interaction between heat transmission and deformation cannot be ignored, 
even in the limiting case of rigid bodies.
 
 We propose ourselves to write a system of equations for temperature
evolution and motion in an elastic bar in the presence of heat transmission.
We will study this problem directly in relativity.

\section{Vibration of an elastic bar}
\subsection{A - Without heat transmission}

We begin with the following problem: How to write in relativity the equation for the adiabatic motion
(without heat transmission) of an elastic bar? An elegant way to do that is as follows.

Let $X$ be the "fixed" coordinates of the bar points and $(x^i,x^4=ct)$ the coordinates of an inertial frame $S$.
The bar motion can be described either using the Lagrange representation $x=x(X,t)$ or using the Euler 
representation $X=X(x,t)$.

Each element $dX$ of the bar has a length $dX_p$ usually subject to deformation with time.
Its proper length at some instant is given by:
 \bc $dX_p =\fr{\fr{\partial x}{\partial X} dX}{\sqrt{1-\beta^2}}~~~~$ with $~~~\beta =\fr{v}{c}~~~$ and
$~~~v=\fr{\partial x}{\partial t}$. \ec
We choose the variable $X$ such that $dX$ is the length of the bar element $dX$ when  not deformed.
The bar deformation ratio is then defined by the variable $s$ given by:
\bc $s = \fr{dX_p}{dX} = s(X,t) = \fr{\fr{\partial x}{\partial X}}{\sqrt{1-\beta^2}}$. \ec
For an elastic body, the pressure $p$ and density $\rho_0$ in a local proper frame must
be functions of both the deformation s and temperature $T$:
\bc $p=p(s,T)~~~$; $~~~\rho_0=\rho_0(s,T)$ \ec
Let $S^*$ be the inertial frame with coordinates $(x^*,x^4*=ct^*)$ that at each instant $t$ travels with each point $X$
having velocity $v=v(X,t)$ in $S$. \\
The components $T^{\alpha^*\beta^*}$ and $T^{\alpha \beta}$of the energy-momentum tensor
of the bar material in the neighborhood of point $X$ in $S^*$ and $S$ are:
\be \begin{array}{ccc}
 T^{\alpha^*\beta^*} =
 \left[ \begin{array}{cc}
p &  0  \\
0 & \rho_0 c^2 \end{array} \right] &  ;  &  T^{\alpha \beta} =
  \left[ \begin{array}{cc}
 \fr{p + \beta^2 \rho_0 c^2}{1 - \beta^2} &
\fr{\beta(p + \rho_0 c^2)  }{1 - \beta^2} \\
\fr{\beta(p + \rho_0 c^2)  }{1 - \beta^2} &
\fr{\beta^2p + \rho_0 c^2  }{1 - \beta^2}
\end{array} \right]
\end{array} \ee

Conservation laws can be expressed, in the one-dimensional case, by the two equations:
\be \partial_\alpha T^{1\alpha} = 0 ~~~~ ; ~~~~ \partial_\alpha T^{4\alpha} = 0 \ee

These conservation laws provide the equations:
\be \begin{array}{lcr} \partial_x \left( \fr{p + \beta^2 \rho_0 c^2}{1 - \beta^2} \right) +
\fr{1}{c} \partial_t \left( \fr{\beta(p + \rho_0 c^2)}{1 - \beta^2} \right) = 0 & ~ ; ~ &
 \partial_x \left(  \fr{\beta(p + \rho_0 c^2)}{1 - \beta^2} \right) +
\fr{1}{c} \partial_t \left( \fr{\beta^2p + \rho_0 c^2}{1 - \beta^2} \right) = 0
\end{array} \ee

Let us now study the motion of an elastic bar. In the adiabatic case we may ignore
the temperature and write $p=p(s)$ and $\rho_0=\rho_0(s)$. As an example, we have the elasticity laws 
of rigid bodies \cite{um}\cite{dois}\cite{tres}:
\be p = \fr{\rho_0^0 c^2}{2} \left[ \fr{1}{s^2} - 1 \right] ~~ ; ~~
\rho_0 = \fr{\rho_0^0}{2} \left[ \fr{1}{s^2} + 1 \right] \ee
Using these laws and doing some calculations, we obtain the following two equations:
\be \left\{  \begin{array}{l}
\fr{\partial X}{\partial x} \left( \fr{\partial^2X}{\partial x^2} - \fr{1}{c^2} \fr{\partial^2X}
{\partial t^2}\right) = 0 \\ \\
 \fr{\partial X}{\partial t} \left( \fr{\partial^2X}{\partial x^2} - \fr{1}{c^2}
\fr{\partial^2X}{\partial t^2} \right) = 0 \end{array}  \right. ,  \ee
where there is a single unknown variable $X=X(x,t)$. The only non-trivial equation is:
\be \fr{\partial^2X}{\partial x^2} - \fr{1}{c^2} \fr{\partial^2X}{\partial t^2} = 0 \ee 
Although similar to the d'Alembert equation (usually written in Lagrange coordinates), this equation is written
in Euler coordinates and, as such, it has a different physical meaning.
\subsection{B - The non-adiabatic case}
In this case, the tensor components $T^{\alpha^*\beta^*}$ and $T^{\alpha \beta}$  are:
\be \begin{array}{lcr}
 T^{\alpha^*\beta^*} =
 \left[ \begin{array}{cc}
p &  \fr{q_0}{c} \\
\fr{q_0}{c} & \rho_0 c^2 \end{array} \right] & ; & 
T^{\alpha \beta} =  \left[ \begin{array}{cc}
\fr{p + \beta^2 \rho_0 c^2 + 2\fr{q_0}{c}\beta}{1 - \beta^2} &
\fr{\beta(p + \rho_0 c^2) + (1 + \beta^2)\fr{q_0}{c}}{1 - \beta^2} \\
\fr{\beta(p + \rho_0 c^2) + (1 + \beta^2)\fr{q_0}{c}}{1 - \beta^2} &
\fr{\beta^2p + \rho_0 c^2 + 2\fr{q_0}{c}\beta}{1 - \beta^2}
\end{array} \right]
\end{array} , \ee
where $q_0$ is the local heat flux in the proper frame $S^*$. The conservation laws give us
the system of equations:
\be \left\{ \begin{array}{c} \partial_x \left( \fr{p + \beta^2 \rho_0 c^2 + 2\fr{q_0}{c}\beta}
{1 - \beta^2} \right) +
\fr{1}{c} \partial_t \left( \fr{\beta(p + \rho_0 c^2) + (1 + \beta^2)\fr{q_0}{c}}
{1 - \beta^2} \right) = 0 \\  \\
 \partial_x \left(  \fr{\beta(p + \rho_0 c^2) + (1 + \beta^2)\fr{q_0}{c}}{1 - \beta^2} \right) +
\fr{1}{c} \partial_t \left( \fr{\beta^2p + \rho_0 c^2 + 2\fr{q_0}{c}\beta}{1 - \beta^2} \right) = 0
\end{array} \right. \ee

These are second-order differential equations in $X=X(x,t)$ and $T=T(x,t)$.
Note that both are invariant under the change from frame coordinates $S$
to another inertial frame $S^{'}$.

To obtain a well defined mathematical system it is necessary to know the constitutive relations
and an appropriate relativistic relation to play the role of Fourier hypothesis:
\be q_0 = - K \frac{\partial T}{\partial X} ~~~~ \left( or ~~~~ q_0 = - K
\frac{\partial T}{\partial x} \right) \ee

Just like Fourier in the XVIII century, what we can do is to adopt some
simple physical hypothesis and look onto the results obtained. If they lead us to
manageable mathematical equations, providing results in agreement with observations
(only expected in Astrophysics), we will be satisfied.

As reasonable assumptions we accept the two  different ones coming from
Fourier's hypothesis (because now $x$ and $X$ are different), and another one built from
the available quantities and respecting the correct physical dimensions:
\bc $ q_0 = - K \left( \frac{\partial T}{\partial X} + R \fr{\gamma_0}{c^2}T \right)$ \ec
where $\gamma_0$ is the acceleration in its proper frame.
\subsection{An application using  the rigid bar elasticity laws}
Let us admit that the elasticity laws obtained for a rigid bar are still valid in non-adiabatic situations.
Using these laws in the non-adiabatic system, presented above, we obtain the equations:
\be  \left\{ \begin{array}{l}
\left( \rho^0_0 c^2  \fr{\partial X}{\partial x} - \fr{2q_0S^2}{c^2} \fr{\partial X}{\partial t} \right)
\left( \fr{\partial^2X}{\partial x^2} - \fr{1}{c^2} \fr{\partial^2X}{\partial t^2} \right)
-2\fr{\partial X}{\partial x} \fr{\partial X}{\partial t} F_x + G F_t  = 0 \\ \\
\left( -\rho^0_0 \fr{\partial X}{\partial t} - \fr{2q_0S^2}{c^2} \fr{\partial X}{\partial x} \right)
\left( \fr{\partial^2X}{\partial x^2} - \fr{1}{c^2} \fr{\partial^2X}{\partial t^2} \right)
-\fr{2}{c^2} \fr{\partial X}{\partial x} \fr{\partial X}{\partial t} F_t + G F_x = 0
\end{array} \right. \ee
where use was made of: $F_x = \fr{\partial}{\partial x} \left( \fr{q_0S^2}{c^2} \right)$ ,
$F_t = \fr{\partial}{\partial t} \left( \fr{q_0S^2}{c^2} \right)$ and $ G = \left[ \left( \fr{\partial X}{\partial x} 
\right)^2 + \fr{1}{c^2} \left( \fr{\partial X}{\partial t} \right)^2 \right]$. \\
The two equations are identical, if the following condition is fulfilled:
\be q_0 = \fr{\rho_0^0 c^3}{2s^2} \ee
We have obtained a new equation of state relating heat flux and deformation.
The two equations degenerate to one:
\be \fr{\partial X}{\partial x} \left( 1-\fr{v}{c}\right) \left[ \fr{\partial^2X}{\partial x^2} - \fr{1}{c^2} \fr{\partial^2X}
{\partial t^2}\right] = 0 \ee
An obvious solution is $v=c$. The other term is identical to the adiabatic case.
\section{Conclusions}

The classical Fourier equation is not a good model for a relativistic variant 
because it studies heat transmission
 in an undeformable bar, while in relativity there are no undeformable bars.
We indicate the way out, showing how to write the system of differential equations including
simultaneously the 3 variables: heat, internal pressure and density.
Meanwhile, we are hostages of the choice of good constitutive relations, relating these 3 variables 
not only with deformation but also with temperature.
These relations should substitute Hooke's law and Fourier's hypothesis.
We present an application using the relativistic adiabatic rigid elastic laws, 
showing agreement with relativistic elasticity.
We have obtained a new constitutive relation between heat flux and deformation.
\footnote{This paper, which was orally presented by one of its authors at the XXVIII Spanish Relativity Meeting, in Oviedo, September 2005,
 has not been included in the conference proceeding because a referee produced the following repport:
"The paper by A. Brotas and C. Fernandez entitled "Heat transfer in theory of relativity" is not acceptable for publication. 
It uses clumsy and obsolete methods for the treatment of the relativistic heat transfer problem whose intrinsic difficulty is 
made worse here by unnecessarily mixing it up with the essentially distinct problem of rigidity.
 The authors seem to know nothing about the vast literature on this subject, starting with the Eckhart theory,
 and proceeding via the textbook treatment of Landau and Lifshitz to more recent work."
It is obvious to us that someone with this opinion will never be able to understand (the resolution of) the problem of heat
 transmission in a vibrating bar, neither in relativity, nor even in classical physics .  
We would like to have Einstein's opinion about this matter.}



\end{document}